# Fully dry PMMA transfer of graphene on *h*-BN using a heating/cooling system

T Uwanno[1], Y. Hattori[1], T Taniguchi[2], K Watanabe[2] and K Nagashio[1,3]
[1] Department of Materials Engineering, The University of Tokyo, Tokyo 113-8656, Japan
[2] National Institute of Materials Science, Ibaraki 305-0044, Japan
[3] PRESTO, Japan Science and Technology Agency (JST), Tokyo 113-8656, Japan
E-mail: nagashio@material.t.u-tokyo.ac.jp

**Abstract**
 The key to achieve high-quality van der Waals heterostructure devices made of stacking various two-dimensional (2D) layered materials lies in the clean interface without bubbles and wrinkles. Although polymethylmethacrylate (PMMA) is generally used as a sacrificial transfer film due to its strong adhesion property, it is always dissolved in the solvent after the transfer, resulting in the unavoidable PMMA residue on the top surface. This makes it difficult to locate clean interface areas. In this work, we present a fully dry PMMA transfer of graphene onto *h*-BN using a heating/cooling system which allows identification of clean interface area for high quality graphene/*h*-BN heterostructure fabrication. The mechanism lies in the utilization of the large difference in thermal expansion coefficients between polymers (PMMA/PDMS) and inorganic materials (graphene/*h*-BN substrate) to mechanically peel off PMMA from graphene by the thermal shrinkage of polymers, leaving no PMMA residue on the graphene surface. This method can be applied to all types of 2D layered materials.

## 1. Introduction

Since the discovery that hexagonal boron nitride (*h*-BN) can serve as an excellent substrate for the fabrication of high quality graphene electronic devices [1], the so-called van der Waals heterostructures made by stacking various two-dimensional (2D) layered materials have been intensively studied [2-28]. One of the critical problems for van der Waals heterostructures is the formation of interfacial bubbles and wrinkles [9,29,30], which are known to deteriorate the performance of the devices [21]. Even in clean room environment, the stacking of graphene and a 2D layered material often results in contaminants trapped at the interface between graphene and the 2D layered material [9]. Therefore, currently, the identification of bubble- and wrinkle-free areas and the subsequent device fabrication at a selected area is essential. Moreover, in the case of encapsulation of graphene with a top 2D layered material, dissolving the sacrificial polymer layer after transferring graphene onto a bottom 2D layered material and the subsequent transfer of the top 2D layered material on graphene resulted in more interfacial bubbles between graphene and the top layered material [21]. Therefore, wet transfer methods in which the sacrificial polymer layers, mainly, polymethylmethacrylate (PMMA), are dissolved [1,3,5-11,13-15,23,25,28] should be avoided.

A fully dry transfer method was achieved by picking up graphene with *h*-BN and depositing the stack on another *h*-BN, encapsulating the graphene [19,21,26]. The range of applications, however, is limited because the transfer capability relies on the ability of *h*-BN to pick up 2D crystal from lower substrate, requiring that the strength of van der Waals force between *h*-BN used to pick up and the target 2D crystal is greater than that between the target 2D crystal and its lower substrate. Recently, van der Waals heterostructures were fabricated in an all-dry manner by applying the transfer printing techniques with polydimethylsiloxane (PDMS) [20,31,32]. The adhesion can be controlled by the peel-off speed due to the viscoelastic nature of PDMS [31,33]. The target object can be picked up from the substrate by peeling PDMS off quickly, while the object attached to PDMS can be transferred onto the target substrate by slow peeling. This simple PDMS dry transfer method was easily reproduced and appears to be promising for the fabrication of van der Waals heterostructures. In this transfer process, target 2D layered materials such as graphene are mechanically exfoliated directly onto PDMS because it is difficult for PDMS to pick up 2D atomic layers that are not sufficiently thick to anchor. In this case, it was realized that monolayer graphene with area as large as 100 μm$^2$ was extremely difficult to obtain by mechanical exfoliation onto PDMS, in contrast to the PMMA-based process. Generally, polymers with high glass transition temperature ($T_g$) exhibit stronger adhesion during the exfoliation due to their high molecular mobilities [33]. The difference in the graphene size can be understood by the comparison of $T_g$ for PDMS (-120°C) and for PMMA (~100°C). Moreover, severe wrinkles indicating lattice defects were introduced in the thick graphite flakes on PDMS due to its softness, again contrary to PMMA (see **supplemental Fig. 1**). Based on these comparisons, PMMA is more promising for the transfer of large-area and high-quality graphene on $h$-BN, but PMMA generally picks up $h$-BN flakes instead of transferring graphene onto $h$-BN, leading to the need to use the conventional wet transfer process so far.

The differences in the thermal expansion coefficients of polymers and inorganic materials, such as graphene, $h$-BN, and SiO$_2$ are quite large (see **supplemental Fig. 2**). In this study, taking advantage of these differences, we demonstrate a fully dry PMMA transfer of graphene onto $h$-BN using the heating/cooling system. The transfer mechanism, bubble formation, and transport characteristics of the fabricated device are discussed.

## 2. PMMA dry transfer

$h$-BN was mechanically exfoliated onto a $n^+$-Si/SiO$_2$ (90 nm) wafer, while graphene was separately exfoliated onto an oxygen plasma treated PMMA/PDMS/glass slide substrate [34,35]. The PMMA/PDMS/glass slide was prepared by directly spin-coating PMMA (PMMA A11, MicroChem) on PDMS (Sylgard 184 Silicone elastomer, Dow Corning). The thickness of PMMA, PDMS and glass slide are ~1 μm, 1.1 mm, and 1 mm, respectively. The root mean square (RMS) for the PMMA/PDMS surface roughness is ~0.7 nm. Confirmation of the flatness in an atomic level at the surface of the thick $h$-BN substrate is the key for obtaining high-quality van der Waals heterostructures. The ~1-nm step measured by atomic force microscopy (AFM) can be detected only by an optical method in which the contrast in bright field mode and differential interference mode is enhanced (**see supplemental Fig. 3**). The layer number of graphene on PMMA was confirmed by Raman spectroscopy. Both substrates were then attached together in a micromanipulator alignment system in the laboratory air condition without any dry gas flow, as shown in **Fig. 1(a)**. Graphene on PMMA and target $h$-BN were aligned under an optical microscope and were mechanically brought into contact by the stepping motor with the resolution of 0.125 μm/pulse and heated. PMMA film was then peeled off from graphene during the cooling of the substrates while PMMA itself remained stick to PDMS/glass slide, as schematically shown in **Fig. 1(b)**. A Peltier module was inserted to control the temperature of the $h$-BN substrate during the transfer process. We note that the heating and cooling can be easily altered without removing the sample by changing the voltage direction.

To investigate the effect of contact temperature on the adhesion of PMMA to $h$-BN on SiO$_2$, the assembly was performed without graphene at different contact temperatures. As shown in **Fig. 2(a)**, at contact temperatures lower than 50°C, non-contact areas were clearly

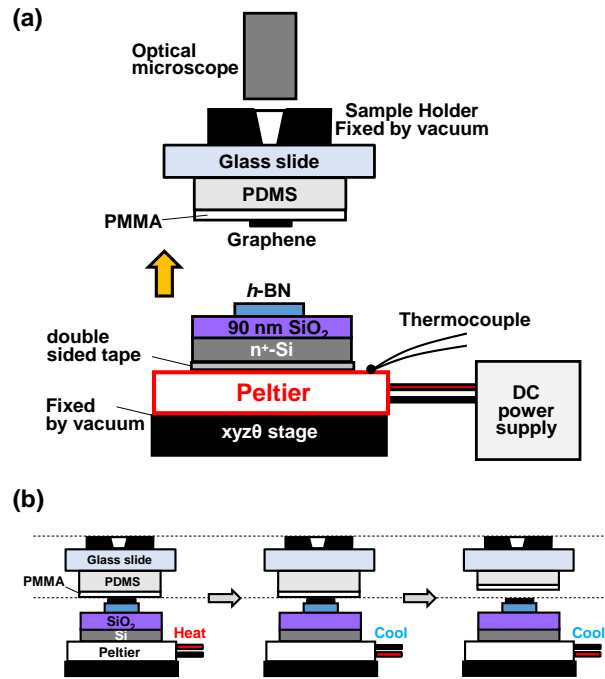

**Figure 1** (a) Schematic illustration of the micromanipulator alignment system. Heating and cooling can be controlled by changing the current direction in the Peltier module. (b) Schematic illustration of the entire transfer process.

identified by the interference fringe formed around 100 nm-thick h-BN flakes. This kind of fringe is not observed for the transfer printing with PDMS [20]. The radius of non-contact area is defined as the length of r in the schematic in **Fig. 2(b)**. These areas decrease when the h-BN substrate is heated above 50°C due to the softening of PMMA and the thermal expansion of PDMS. This softening temperature is consistent with $T_g$ of PMMA measured by the differential thermal analysis (see **supplemental Fig. 4**). The thermal expansion of PDMS enables the softened PMMA to completely conform to the shape of the thick h-BN flakes at 90°C, thus ensuring the firm contact between graphene and target h-BN.

Based on this series of study, it is realized that once graphene and h-BN are brought into contact, their adhesion is strong enough to either cause h-BN to be picked up by graphene on PMMA or to cause graphene to be transferred to h-BN on Si/SiO2. To successfully transfer graphene onto h-BN, h-BN must not be picked up by PMMA. After the contact at two different temperatures of $T_{cont}$ = 55 and 90°C, the effect of the peel-off temperature on the fraction of the h-BN flakes picked up by PMMA was investigated. As shown in **Fig. 2(c)**, a higher fraction of h-BN flakes was picked up by PMMA at $T_{cont}$ = 90°C than at $T_{cont}$ = 55°C for all peel-off temperatures; this is due to firm contact around the edge of h-BN flakes (anchoring effect). These data were obtained by counting the number of h-BN flakes on the Si/SiO2 substrate before and after the transfer in the selected area (~450 μm×350 μm). It was found that the contact temperature of approximately 55°C and the peel-off temperature of approximately 15°C are optimal due to the lower fraction of h-BN picked up by PMMA as well as the ensured contact between graphene and h-BN. **Fig. 2(d)** shows an AFM image of successfully transferred graphene on h-BN. For these conditions, the success rate is greater than 80 %. The holding time at 55 °C is important in this dry transfer method. If the holding time at 55 °C is too long, e.g., 5 min, the adhesion between PMMA and SiO2 becomes too strong, causing PMMA to detach from PDMS. Even for the holding time of ~3 min, h-BN is firmly anchored by PMMA, causing h-BN to be

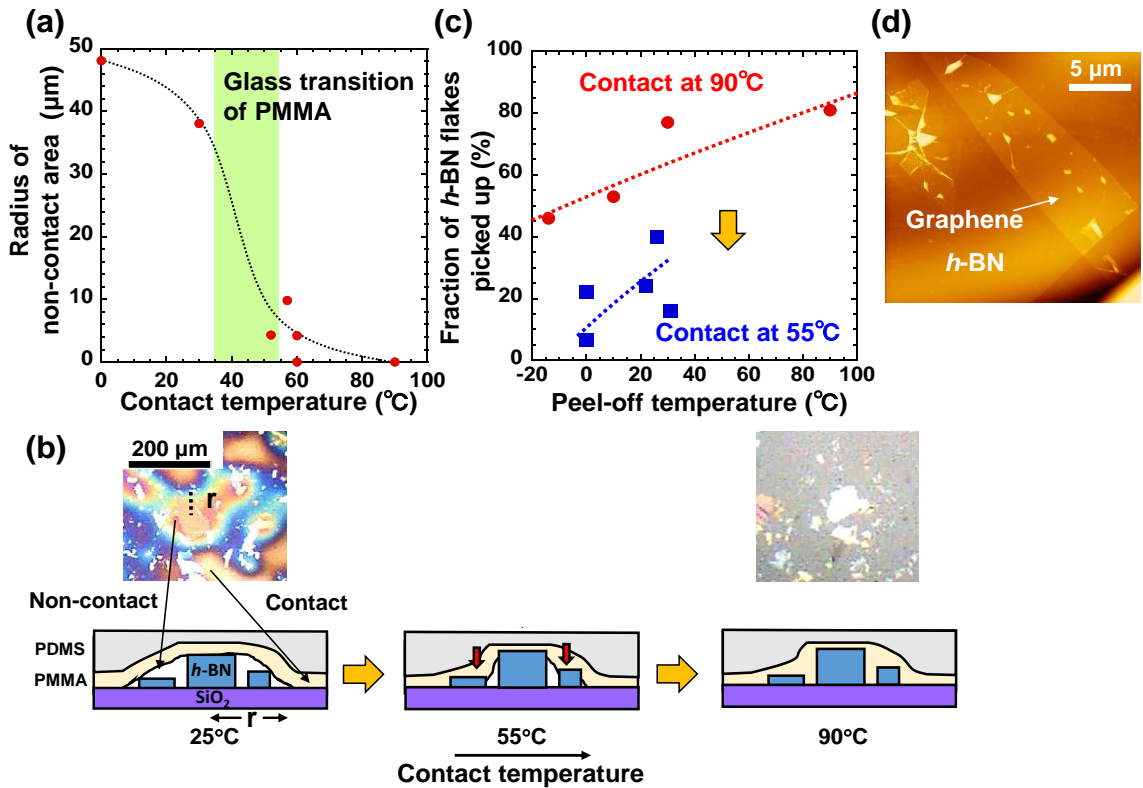

**Figure 2** (a) Relationship between radius of non-contact area and contact temperature. (b) Optical microscope images and schematics of the contact between PMMA/PDMS and h-BN/SiO2. The definition of radius of non-contact area is shown as r in the schematic. (c) Fraction of h-BN flakes picked up by PMMA as a function of peel-off temperatures at different contact temperatures. (d) AFM image of graphene successfully dry-transferred onto h-BN.

picked up by PMMA. Moreover, the peel-off speed should be slow enough (~ 80 μm/s), otherwise *h*-BN can be picked up easily, due to anchoring effect. Sudden cooling after heating usually results in relatively fast peel-off speed.

We now discuss the mechanism of the dry PMMA transfer. In general, the adhesion between PMMA and inorganic materials would be mainly controlled by van der Waals force represented by London dispersion force, which increases with decreasing temperature [36]. Therefore, this dry transfer cannot be explained by the temperature dependence of van der Waals force. As indicated in introduction, the difference in the thermal expansion coefficient values for polymers and inorganic materials is the key in this transfer. After graphene and *h*-BN were brought into contact at 55°C, PMMA can be automatically peeled off at 15°C without moving the stage down, suggesting that the thermal shrinkage of PDMS is the origin for this movement, as schematically shown in **Fig. 1(b)**. To quantitatively estimate the contribution of the thermal expansion of PDMS, a PDMS/glass slide and a bare Si/SiO$_2$ substrate were attached together by the micromanipulator (see **supplemental Fig. 5**). Heating PDMS from 15 to 55°C led to an increase in the contact area between PDMS and SiO$_2$. On the other hand, the same increased contact area can also be obtained by moving Si/SiO$_2$ up against PDMS. In this way, the change of PDMS thickness can be measured by changing the height of the bottom stage with the stepping motor. The amount of PDMS deformation estimated from the thermal expansion during the temperature change from 15 to 55°C was $1.07 \times 10^{-2}$ mm, which was roughly consistent with the change of the stage height ($1.18 \times 10^{-2}$ mm). This simple estimation supports the hypothesis that the fully dry PMMA transfer is controlled by thermal expansion. It should be noted that the dry transfer using the thermal expansion allows much smoother peeling than that using the stepping motor.

The advantage of this method is that it is easy to select the clean interface area without bubbles and wrinkles for the FET channel, as

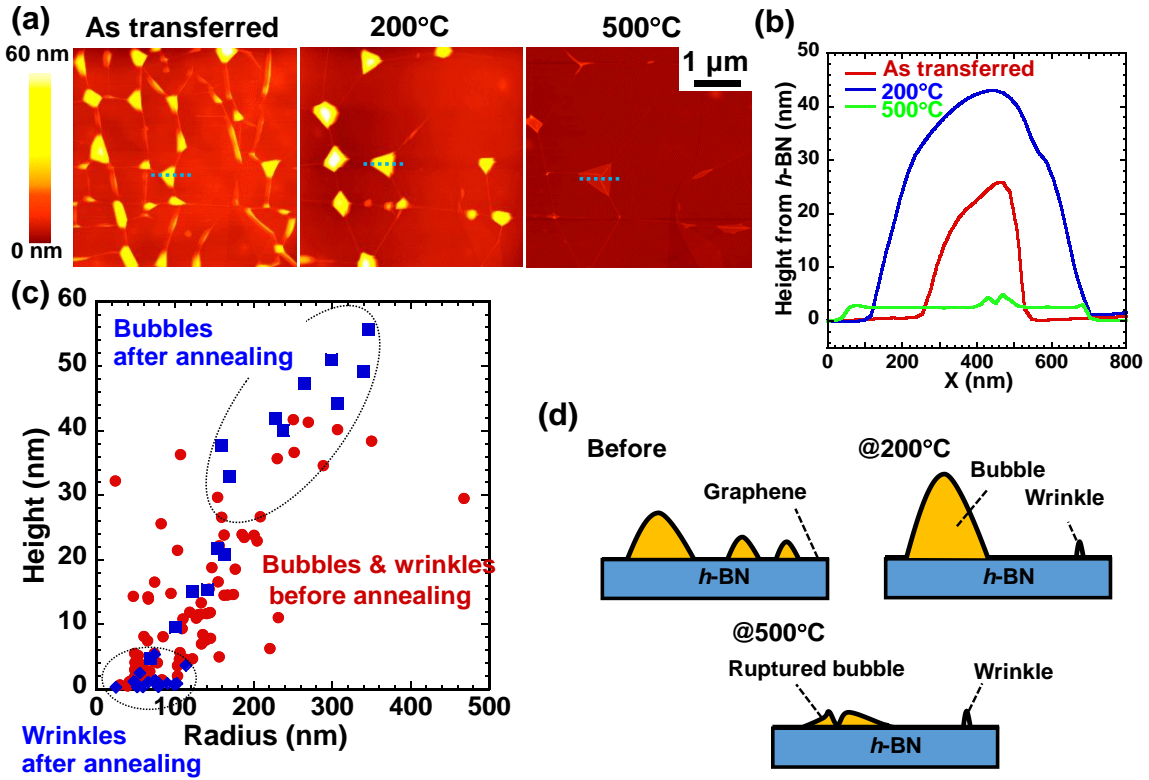

**Figure 3** (a) AFM images of graphene transferred to *h*-BN before and after annealing at different temperatures. (b) Height profile of bubbles along the dotted lines in (a). (c) Height and radius of bubbles and wrinkles before and after annealing at 200°C estimated from (a). Red circles (●), blue squares (■), and blue diamonds (♦) indicate the bubble and wrinkle before annealing, bubble after annealing, and wrinkle after annealing, respectively. (d) Schematic illustrations of bubbles and wrinkles before and after annealing at 200°C and 500°C.

clearly seen in **Fig. 2(d)**, due to PMMA being peeled off from graphene mechanically, leaving no PMMA residue on graphene surface. On the contrary, in previously reported wet methods [1,3,5-11,13-15,23,25,28], polymer films are dissolved after the transfer, making it more difficult to find the clean interface area, especially the identification of wrinkles-free areas. This is because polymer residue on graphene has almost the same height as the wrinkles even after annealing. This issue on the visibility of wrinkles becomes more critical in case of encapsulating graphene with $h$-BN [19], because having top $h$-BN layer with thickness of several nanometers results in less visibility of bubbles and wrinkles due to the thickness of top $h$-BN. Of course, annealing in Ar/$O_2$ at 500°C was reported to be an efficient way to remove PMMA residue on h-BN [37], but graphene outside of $h$-BN encapsulation would be damaged. We have already achieved the fully dry encapsulation of graphene by $h$-BN using this dry transfer method. Finally, it should be noted that this dry transfer possesses the wide applicability to all types of 2D layered materials, because the thermal expansion coefficients for inorganic 2D materials are similar.

## 3. Bubbles at graphene/$h$-BN interface

Unlike graphene transferred onto $h$-BN by dissolving PMMA, wrinkles and bubbles can be clearly seen, while no PMMA residue was observed at the surface of graphene transferred onto $h$-BN by dry PMMA transfer, as shown in **Fig. 3(a)**. After the stack was annealed in Ar/$H_2$ gas at 200°C for 3 hours, small bubbles aggregated into larger bubbles to reduce the total surface energy; this is the so-called Ostwald ripening. The height and radius of the bubbles and wrinkles before and after annealing were measured and are plotted in **Fig. 3(c)**. The number of bubbles decreased after annealing, while the size and height of bubbles increased. After annealing, the height and width of the wrinkles were roughly the same throughout the sample. The clean area increased. Unfortunately, the total volume of bubbles appears to be unchanged for this sample. This behavior is summarized schematically in **Fig. 3(d)**.

To investigate the possibility for extending the clean interface area, the annealing was carried out at more elevated temperatures (300ºC, 400ºC, and 500ºC) for 3 hours. Annealing graphene/$h$-BN at above 400ºC resulted in the rupture of the bubbles (not shown here). There are two main possible origins of the rupture: the reaction between the graphene and the oxygen trapped within the bubbles and thermal $sp^2$ bond breaking due to large strain induced by high curvature of the bubble. The presence of oxygen in the bubbles was investigated by Auger electron spectroscopy using another sample where aggregated bubbles were still intact. The bubble areas show a higher oxygen content than the bubble-free areas on graphene and on $h$-BN. This is consistent with the previous $O_2$ annealing experiment where the D band is clearly observed at 400 °C for $O_2$ annealing [38]. The rigorous control of transfer environment is quite important. While further annealing at 500ºC appears to cause the ruptured bubbles to shrink as the content escaped from the bubbles (**Figs. 3(a) and (b)**), the wrinkles around the bubbles did not disappear. The high temperature annealing did not further increase the clean interface area of graphene/$h$-BN for device fabrication.

## 4. Demonstration of high mobility of graphene/$h$-BN

After annealing graphene/$h$-BN stack in Ar/$H_2$ gas at 200°C for 3 hours, the bubble region was removed by oxygen plasma and field-effect transistor devices were fabricated at the clean interface region using the conventional electron lithography technique. Finally, the device was annealed in Ar/$H_2$ gas at 300°C for 3 hours. Transport characteristics in four-probe graphene device on $h$-BN are shown in **Fig. 4(a)**. Compared with the graphene device on $SiO_2$, the resistivity curve is much sharper and the resistivity at the Dirac point is much higher. The thickness of $h$-BN is 47 nm as measured by AFM. The carrier density is estimated using the dielectric constants of 4 and 3.9 for $h$-BN and $SiO_2$, respectively. The dielectric constant of top gate $h$-BN was estimated to be 4 from the Dirac point shift in the electrical measurement for $h$-BN/graphene/$SiO_2$ dual gate device [39]. Therefore, the same value was used in this carrier density calculation. At 20K, the carrier mobility is ~20,000 $cm^2V^{-1}s^{-1}$ at the carrier density of $2\times10^{11}$ $cm^{-2}$.

The temperature dependence of the resistivity at the Dirac point can be used as the measure of the charged impurity density [40,41].

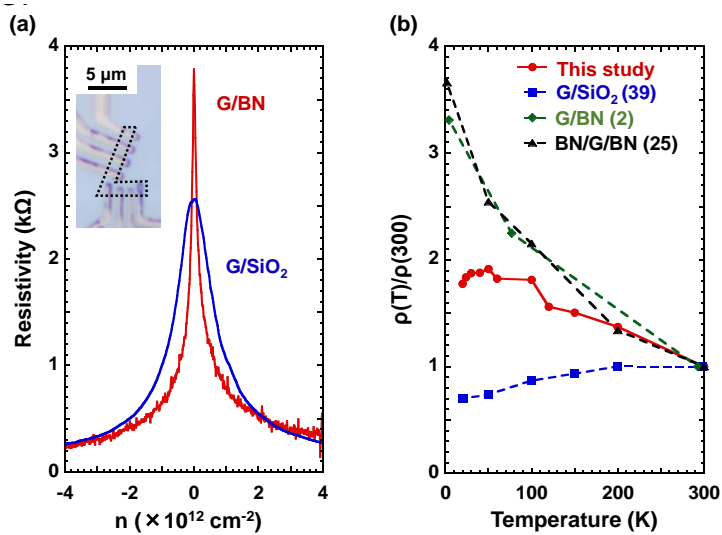

**Figure 4** (a) Resistivity as a function of carrier density at 20K. (b) Temperature dependence of normalized resistivity at the Dirac point. Literature data is also included for comparison.

The resistivity at the Dirac point for graphene on Si/SiO$_2$ substrate shows almost no temperature dependence, because the carrier density induced externally by the charged impurities is larger than the density of thermally excited carriers around the Dirac point due to the linear dispersion [42]. On the other hand, the resistivity at the Dirac point for graphene on *h*-BN increases with decreasing temperature, as shown in **Fig. 4(b)**, indicating less charged impurities at the graphene/*h*-BN interface. These data suggest the validity of fully dry PMMA transfer of graphene on *h*-BN.

## 5. Conclusions

Utilizing the large difference between the thermal expansion coefficient values of PMMA/PDMS and graphene/*h*-BN substrate, fully dry transfer of graphene onto *h*-BN was achieved by using a heating/cooling system, enabling the identification of clean interface areas for the device fabrication. Annealing the stack in Ar/H$_2$ gas at 200 °C caused small bubbles to aggregate into larger bubbles, widening the clean area. Further annealing at higher temperatures caused the bubbles to rupture and shrink, but did not increase the clean areas due to the continued presence of wrinkles. Transport data for the devices fabricated by this method shows superior characteristics that cannot be achieved in graphene/SiO$_2$ devices, suggesting the validity of this transfer method.

**Acknowledgements:** We are grateful to Covalent Materials for kindly providing us Kish graphite. This research was partly supported by Grants-in-Aid for Scientific Research on Innovative Areas by the Ministry of Education, Culture, Sports, Science and Technology of Japan.

**Supplemental note:**

# Fully dry PMMA transfer of graphene on *h*-BN using heating/cooling system

T Uwanno[1], T Taniguchi[2], K Watanabe[2] and K Nagashio[1,3]

[1] Department of Materials Engineering, The University of Tokyo, Tokyo 113-8656, Japan

[2] National Institute of Materials Science, Ibaraki 305-0044, Japan

[3] PRESTO, Japan Science and Technology Agency (JST), Tokyo 113-8656, Japan

**E-mail:** nagashio@material.t.u-tokyo.ac.jp

**On PDMS Gel-Pak PF-3-X4**

(a)
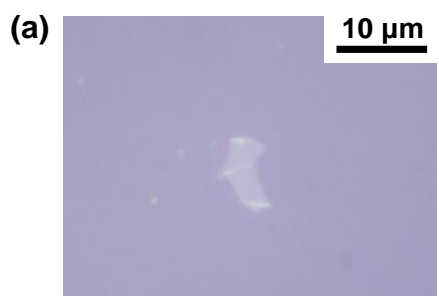
10 µm

(c) BF | DIC
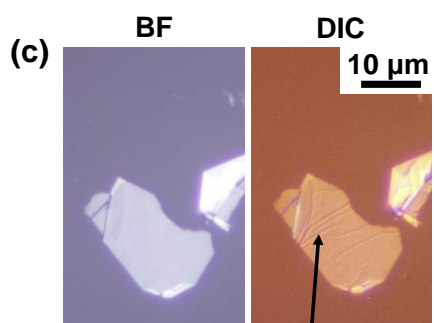
10 µm
**Wrinkles**

**On PMMA/PDMS MICRO·CHEM**
PMMA A11 $M_w$ 495,000

(b)
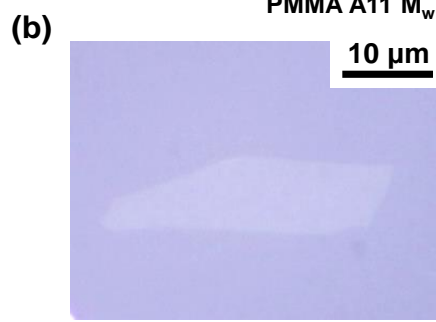
10 µm

(d) BF | DIC
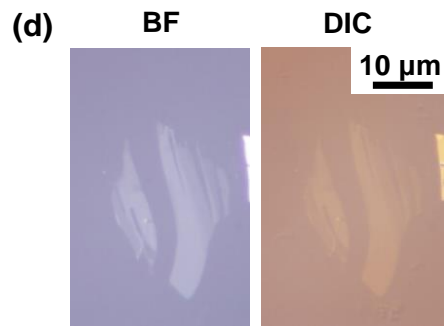
10 µm

**Supplemental Figure S1** (a) Optical image of graphene exfoliated onto PDMS. (b) Optical image of graphene exfoliated onto PMMA/PDMS. (c) Optical image of thick graphite exfoliated onto PDMS in bright field mode (BF) and differential interference mode (DIC). (d) Optical image of graphite exfoliated onto PMMA/PDMS in bright field mode (BF) and differential interference mode (DIC).

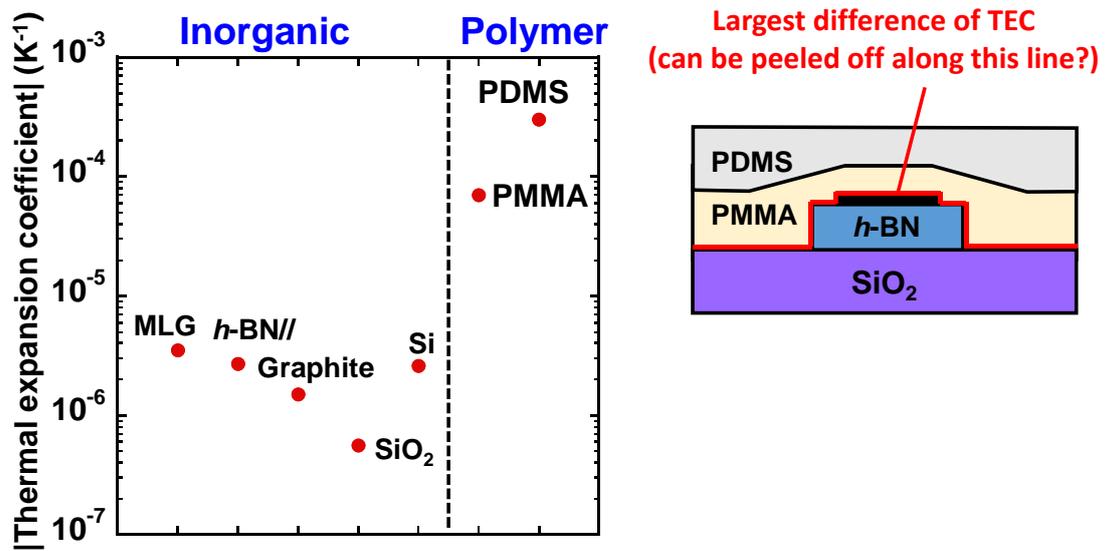

**Supplemental Figure S2** Absolute value of linear thermal expansion coefficient of monolayer graphene, $h$-BN (c axis), graphite, $SiO_2$, Si, PMMA, and PDMS.

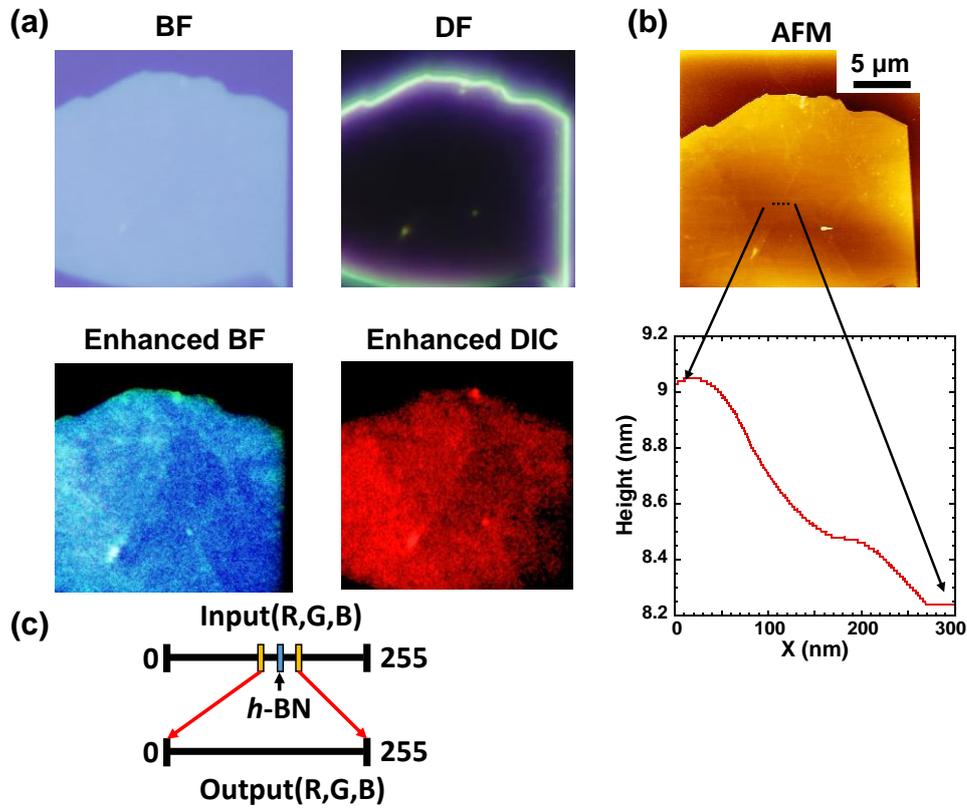

**Supplemental Figure S3** (a) Optical images of ~10 nm thick *h*-BN in bright field mode, dark field mode, enhanced bright field mode, enhanced differential interference mode. (b) AFM image of *h*-BN in (a). Bottom figure is the height profile for the step with ~1 nm on ~10 nm thick *h*-BN surface. (c) Schematic of contrast enhancing method. The input range of R, G, B value are set to the range around the value of *h*-BN, while the output range is maximized. According to contrast calculation based on Fresnel equation, the optical contrast of *h*-BN on Si/SiO$_2$ (90 nm) is highest at wavelength 750 nm. Therefore, for DIC mode in which the red color is selected by adjusting the DIC prism, only the input range of R is set to the range around the value of *h*-BN, while the range of input G and B is set to the value far larger than *h*-BN in order to observe only in red light.

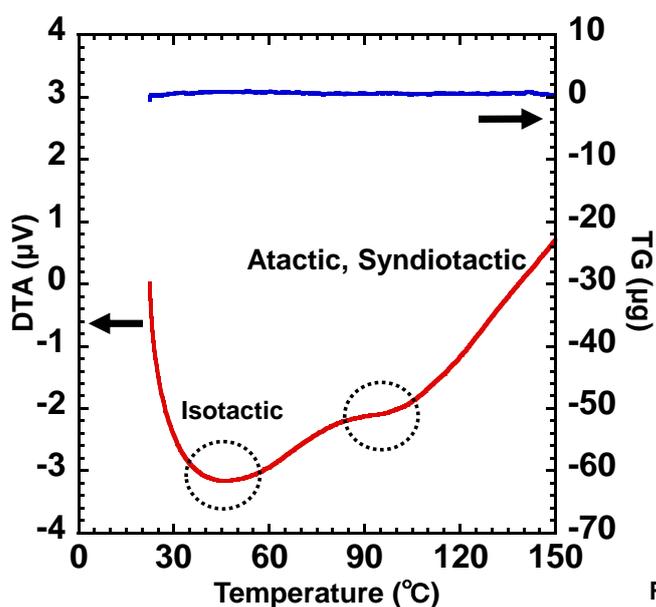
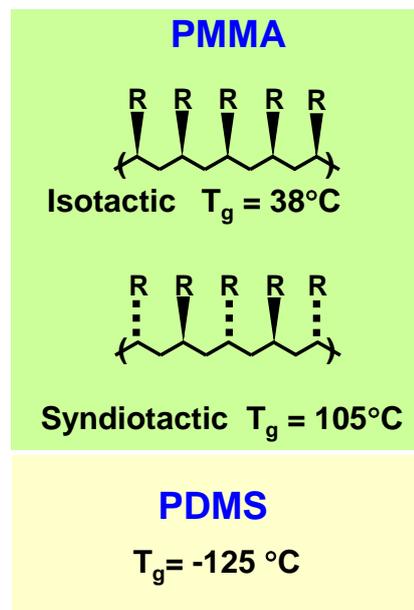

**Supplemental Figure S4** Thermogravimetric/ Differential thermal analysis (TG/DTA) of PMMA was carried out. The measurement was iteratively repeated to evaporate anisole (solvent for PMMA). The DTA data of PMMA showed two endothermic peaks without mass change (no change in TG), indicating the two different glass transition temperatures ($T_g$ = ~50 and ~100°C). Although exact contents are not known because PMMA was purchased from MircoChem, both isotactic structure and syndiotactic structure may be present in the present PMMA.

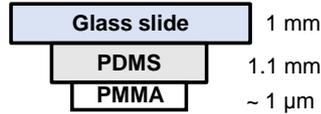

$$\frac{\Delta L}{L} = \alpha_L \Delta T$$

*ΔL*: Change of length
*L*: Original length
*α_L*: Linear thermal expansion coefficient
*ΔT*: Change of temperature

☑ Change of thickness when heated from 15°C to 55°C

| Material | $\alpha_L$ ( × 10⁻⁶ K⁻¹) |
|---|---|
| SiO$_2$ | 0.56 |
| PMMA | 50 - 90 |
| PDMS | 300 |

Glass slide — 1 mm
PDMS — 1.1 mm
PMMA — ~ 1 μm

**Change of thickness**
Glass slide  $1.80 \times 10^{-8}$ m
PMMA  $2.24 \times 10^{-9}$ m
PDMS  $1.07 \times 10^{-5}$ m

We should consider the change of thickness only for PDMS.

☑ Simple estimation only using PDMS

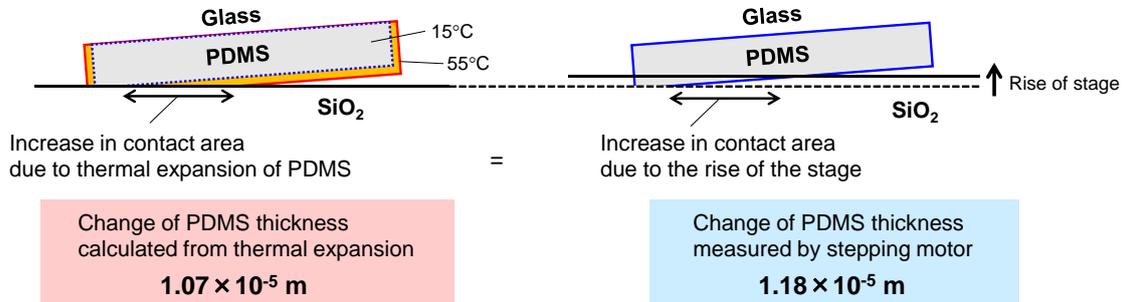

● 1st exp: Temperature change from 15 to 55°C
● 2nd exp: Stage move up by stepping motor @ 15°C

Increase in contact area due to thermal expansion of PDMS = Increase in contact area due to the rise of the stage

Change of PDMS thickness calculated from thermal expansion
**$1.07 \times 10^{-5}$ m**

Change of PDMS thickness measured by stepping motor
**$1.18 \times 10^{-5}$ m**

**Thickness change is due to thermal expansion.**

**Supplemental Figure S5**  The change of thickness can be estimated by the equation shown in the figure. In this estimation, the thickness of polymers as well as the linear thermal expansion coefficient is important. The change of thickness of PMMA and glass slide during the temperature change from 15 to 55°C are negligible compared to PDMS.

To estimate the contribution of the thermal expansion of PDMS quantitatively, PDMS/glass slide and bare Si/SiO$_2$ substrate were attached together by micromanipulator. Heating PDMS from 15 to 55°C caused the contact area between PDMS and SiO$_2$ to increase. On the other hand, the same increased contact area can also be obtained by moving Si/SiO$_2$ up against PDMS. This way, the change of PDMS thickness can be measured by the change of the height of the bottom stage with the resolution of 0.125 μm/pulse. The amount of PDMS deformation estimated from the thermal expansion during the temperature change from 15 to 55°C was $1.07 \times 10^{-2}$ mm, which was roughly consistent with the change of the stage height ($1.18 \times 10^{-2}$ mm). This simple estimation supports the idea that fully dry PMMA transfer is controlled by the thermal expansion.